\documentclass[12pt]{article}
\usepackage{amssymb,bbold,euscript}

\begin{document}

\vspace{-10mm}
\begin{flushright}
\hfill{UPR-938-T\ \ HU-EP-01/39 \ \ NSF-ITP-02-12\ \ RUNHETC-2001-30}\\
\hfill{hep-th/0201272}

\end{flushright}

\vspace{10pt}

\begin{center}{ \Large{\bf
Bent BPS domain walls of D=5 N=2  gauged \\[2mm]
supergravity  coupled to hypermultiplets
}}

\vspace{15pt}

{Klaus Behrndt$^a\,
^d$\footnote{E-mail:behrndt@physik.hu-berlin.de}
 and Mirjam Cveti\v c$^b \, ^c\, ^d$
\footnote{E-mail:cvetic@cvetic.hep.upenn.edu}
 }
\vspace{15pt}

{\it $^a$ Humboldt Universit\"at zu Berlin,
Institut f\"ur Physik,\\
Invalidenstrasse 110, 10115 Berlin,
 Germany\\
$^b$ Department of Physics and Astronomy, \\
University of Pennsylvania, Philadelphia, PA 19104-6396, USA
\\
$^c$ Department of Physics, Rutgers University,\\
 Piscataway,  NJ 08855, USA
 \\
 $^d$ Institute for Theoretical Physics, \\
 University of California, Santa Barbara, CA 93106-4030\\
} \vspace{10pt}

\underline{ABSTRACT}
\end{center}

Within D=5 N=2 gauged supergravity coupled to hypermultiplets we
derive consistency conditions for BPS domain walls with constant
negative curvature on the wall.  For such wall solutions to
exist, the covariant derivative of the projector, governing the
constraint on the Killing spinor, has to be non-zero and
proportional to the cosmological constant on the domain walls. We
also prove that in this case solutions of the Killing spinor
equations are indeed solutions of the equations of motion.  We
present explicit, analytically solved examples of such domain
walls, employing the universal hypermultiplet fields. These
examples involve the running of two scalar fields and the
space-time in the transverse direction that is cut off at a
critical distance, governed by the magnitude of the negative
cosmological constant on the wall.

\newpage
\newcommand{\be}[3]{\begin{equation}  \label{#1#2#3}}
\newcommand{\bib}[3]{\bibitem{#1#2#3}}
\newcommand{\ee}{\end{equation}}
\newcommand{\ba}{\begin{array}}
\newcommand{\ea}{\end{array}}
\newcommand{\p}{\partial}

\let\huge=\Large
\let\LARGE=\Large
\let\Large=\large



\section{Introduction}

The study of BPS domain wall solutions in supergravity theory was
initiated in \cite{cgr} where the first examples of such solutions
within D=4 N=1 minimal supergravity coupled to chiral superfields were
found. These examples correspond to static, flat domain walls where
the neutral scalar component of a chiral superfield interpolates
between isolated supersymmetric extrema of the matter potential with
non-positive cosmological constant.  Such solutions were further
generalized \cite{c} to examples of BPS domain walls which couple to
the linear supermultiplet, leading to walls interpolating between
isolated supersymmetric vacua with the varying dilaton field,
corresponding to a scalar component of the linear supermultiplet. (For
a review and generalizations to non-BPS domain wall configurations,
see \cite{cs}.)

The study of BPS domain wall configurations found new implications
within AdS/CFT correspondence. There remains an on-going effort to
elucidate D=4 N=1 super Yang-Mills theories as a dual
description in terms of gauged supergravity in five dimensions or
its decompactification to 10 or 11 dimensions. {From} the
five-dimensional perspective this corresponds to finding domain wall
solutions of the Killing spinor equations. One direction was to
consider (deformed) sphere compactification of type IIB string
theory, which can be seen as gauging of N=8 supergravity. (For
the study of  consistent non-linear Kaluza-Klein  sphere
compactifications see, e.g., \cite{clp} and references therein.)
In this approach the domain wall solutions parameterize the
renormalization group (RG) flows that  preserves only four super
charges of the N=1 dual field theory.  By now there is a plethora of
RG flow examples, with the best known example discussed in
\cite{250}.  In a complementary approach one can address these
questions directly within N=2 gauged supergravity, which
contains as a subclass the models that can be understood as from
N=8 supergravity. For the above flow this was done in
\cite{080}.

Another important role played by BPS domain wall is that gravity could
be trapped on the wall, providing the warp factor, describing the
asymptotic AdS space-times, falls-off on both sides of the wall
\cite{rs}. Within N=1 D=4 minimal supergravity such example have been
established \cite{cgr,cg} when the corresponding vacua of the matter
potential are minima of the matter potential and the superpotential
changes sign on either side of the wall. (The thin wall limit of such
walls was analyzed in \cite{bcy}.) On the other hand within D=5 gauged
supergravities, (such supergravities are believed to arise as
consistent compactifications of M- or string theory), the existence of
such gravity trapping, smooth configurations, remains elusive.  While
by now there is a rich plethora of domain wall solutions describing RG
flows in dual field theories, on the other hand within D=5 N=2 gauged
supergravities coupled to vector and tensor supermultiplets, so-called
no-go theorems for the existence of smooth BPS domain walls that could
trap gravity were established \cite{150,140}. If one includes
non-trivial hypermultiplets the flow towards the infra-red can be
regular, but there are no truly infra-red (IR) critical points -- only
saddle points with some IR directions are possible (at least as long
as the scalar manifold is homogeneous) \cite{al/co}. See however,
recent work that establishes the existence of the smooth gravity
trapping solution realized by using non-homogeneous scalar manifolds
\cite{270}.

Another interesting direction constitutes a construction of BPS domain
walls that are not flat, but have a constant negative curvature,
i.e. the space-time on the wall is AdS.  [The thin AdS vacuum domain
walls, with the negative (as well as positive) curvature were first
studied in D=4 in \cite{cgs}. For generalizations to D-dimensions,
see, e.g. \cite{060,cw}.] It turns out that such configurations may
provide a background where gravity is ``locally localized''
\cite{210}, providing the warp factors fall-off in the sufficient
vicinity on either side of the wall. However, the explicit realization
of such configurations within D=5 gauged supergravity may prove to be
a difficult task.  On the other hand, explicit examples of bent BPS
domain walls, even though their warp factors may not provide for the
localization of gravity, remain to be of interest from the AdS/CFT
correspondence.  Establishing explicitly the existence of such
configurations within D=5 N=2 gauged supergravity is the main purpose
of this paper.

The main difficulty in establishing a more general classification of
BPS domain wall configurations within D=5 N=2 gauged supergravity is
due to the fact N=2 gauged supergravity with general matter couplings
has been worked only recently \cite{090,170}. The hypermultiplet
sector is very much similar to the D=4 N=2 gauged supergravity case
which was known for some time (see \cite{100} for a review).
So far only very few models could be discussed explicitly
\cite{030, 080, be/he}, but unfortunately most the explicit
examples are singular (note, however,   regular flows can be
constructed using the model of \cite{270}).

As far as the existence of bent BPS domain walls in gauged
supergravity goes, within D=4 N=2 gauged supergravity, their
existence and explicit examples were studied in \cite{010}.
Within D=5, the discussion of bent AdS walls is given in
\cite{070,040,050}, and examples with trivial scalars was given
in \cite{210}.

In this paper we advance the study of bent BPS domain walls in several
ways. In D=5 N=2 gauged supergravity theory coupled to a general set
of hypermultiplets we derive first order Killing spinor equations for
the metric (eq. (\ref{532})) , scalar fields (eq.(\ref{620})) and the
$SU(2)$ valued projector $\Theta\equiv \Theta^x\sigma^x$ (where
$\sigma^x$-Pauli matrices, and $\Theta^x$-``phase factors''
($x=1,2,3$)) (eq. (\ref{543})). This projector governs the generalized
constraint on the Killing spinor. In particular, we find that the
generalization to the curved BPS walls involves a nontrivial projector
whose covariant derivative (in the transverse direction of the wall)
is non-zero and proportional to the cosmological constant on the
domain wall (see eqs.  (\ref{829}), (\ref{736})).  A detailed
derivation of the Killing spinor equations and constraint for these
bent BPS walls is given in Section 2.

We also prove that the first order system of the Killing spinor
equations indeed satisfy the equations of motion, and thus genuinely
represents the BPS solutions of the theory. The key ingredient role in
the proof is played by the nontrivial projector. The proof is given in
Section 3.

In Section 4 we present two explicit examples of such bent walls,
employing the fields of the universal hypermultiplets. The result can
be represented in the analytic form, and involves two scalar fields.
The generic property of these space-times is that at a critical value
of the transverse coordinate, governed by the cosmological constant on
the wall, the space-time is cut off.


\section{Solving the Killing spinor equations}


N=2 supergravity in five dimensions has eight supercharges and matter
fields enter vector, tensor or hypermultiplets. We consider the case
where all vector and tensor multiplets are trivial and all scalars are
part of hypermultiplets. In addition, the gravity multiplet has besides
the graviton one graviphoton as bosonic fields. Since this model contains
only one Abelian vector (the graviphoton), we can gauge only an Abelian
symmetry and the $SU(2)$ R-symmetry of N=2 supergravity will be
broken to an Abelian subgroup. Due to supersymmetry the hyper scalars,
which we will denote by $q^u$, have to parameterize a quaternionic space
{$\cal M$} \cite{160} with the metric $h_{uv}$ and without making to much
restrictions we can assume the existence of a number of isometries of
${\cal M}$ parameterized by Killing vectors. Since {$\cal M$} is
quaternionic, the holonomy group is contained in $SU(2) \times Sp(n)$ and
there is a triplet of covariantly constant K\"ahler 2-forms $\Omega^x$,
where $x = 1,2,3$ is the $SU(2)$ index; for a recent nice summary of
quaternionic geometry we refer to the appendix of \cite{110}.  Using
these 2-forms one introduces for a given Killing vector $k$ a triplet of
Killing prepotentials $P^x$ as follows
\be826
P^x = \Omega^{x\, uv} (\partial_u k_{v}) \equiv \Omega_{rs}^x\,
    h^{ru} \, h^{sv} \, (\partial_u k_v) \ ,
\ee
which  solve the following  equation:
\be284
\Omega^x_{uv} k^v = - (\nabla_u P)^x \equiv - (\partial_u P^x + \epsilon^{xyz}
 \omega^y_u P^z )\ ,
\ee
where $\omega^x_u$ is the $SU(2)$ part of the spin connection,
$(\omega_u)_i^{\ j} \equiv i \, \omega^x_u (\sigma^x)_i^{\
j}$.

Before we can start to investigate bent domain wall solutions we have to
discuss the gauge fields.  Domain walls are codimension one objects and
should not be charged with respect to vector fields. Nevertheless the
decoupling of the vector fields is a subtle point: as a consequence of
the gauging, the scalar fields correspond to charged matter and
basically represent sources for gauge fields.  In fact, gauging the
isometry $q^u \simeq q^u + k^u$ yields the covariant derivative $D q^u =
d q^u + k^u A$ and the source current becomes $J \sim k_u d q^u$. As we
will see in eq.\ (\ref{562}) the scalar flow becomes perpendicular to
the Killing direction and therefore the charged scalars can remain
constant.  Hence, the correponding current will vanish and the gauge
fields can be consistently decoupled.

For a BPS configuration the fermionic supersymmetry variations, which
have been derived for general couplings in \cite{090}, have to vanish. In
our notation they are given by
\be410
\ba{l}
\delta \psi_{\mu i} = D_{\mu} \epsilon_i - {i \over 2} \Gamma_{\mu} S_{ij}
    \epsilon^j \ , \\
\delta \zeta_{\alpha} = V^i_{u \alpha} \Big[ {i \over 2}
    \Gamma^{\mu} \partial_{\mu} q^u  +i\, {3 \over 2}  \, k^u \Big]
    \epsilon_i\ ,
\ea
\ee
where
\be420
\ba{l}
D_{\mu} \epsilon_i = (\partial_{\mu} + {1 \over 4} \omega_{\mu}^{ab}
    \Gamma_{ab}) \epsilon_i - i (Q_{\mu})_i^{\ j} \epsilon_j \ , \\
(Q_{\mu})_i^{\ j} = {i \over 2} \, \partial_{\mu}q^u \omega_u^x
    (\sigma^x)_i^{\ j} \ , \\
S_{ij} = i \, P^x (\sigma^x)_i^{\ k} \epsilon_{jk}\ ,
\ea
\ee
where $\sigma^x$ are the Pauli matrices.

For a domain wall solution we make the metric Ansatz
\be450
ds^2 = e^{2A(z)} \hat e^m \hat e^m + dz^2\ ,
\ee
where $\hat e^m \hat e^m = d\hat s^2$ is the line element of the wall
of constant negative curvature ($\hat R_{m}^{\ n} = - 3 \lambda^2
\delta_{m}^{\ n}$). We should take a four-dimensional AdS space, which
ensures that the wall curvature does not give rise to further breaking
of supersymmetry and the Killing spinor satisfies the equation
\be481
\hat D_m \epsilon_i = {\lambda \over 2} \, Q^x \sigma^x \Gamma_m \epsilon_i
\qquad {\rm with}: \quad |\, Q^x\, |^2 = 1  \ ,
\ee
where we introduced a general SU(2) phase (with $\hat D_m Q^x = 0$) to
solve the equations below. For the supersymmetry projector we make the
following general Ansatz (we have summarized our convention in an
Appendix):
\be490
\epsilon_i = - \Gamma_5 \Theta_i^{\ j} \epsilon_j\ ,
\ee
with $\Gamma_5 = \Gamma_z$. This is a consistent projector if
$\Theta_i^{\ k} \Theta_k^{\ j} = \delta_i^{\ j}$ and therefore one
writes $\Theta_i^{\ j}$ as
\be510
\Theta = \Theta^x \sigma^x \qquad {\rm with} \quad |\Theta^x |^2 = 1 \ .
\ee
Assuming that the warp factor as well as the scalars depend only
on the radial coordinate ($A = A(z), q^u = q^u(z)$ with $\dot A \equiv
\partial_z A$)  the gravitino variation $\delta \psi_{m i}$
($m = 1\ldots 4$) gives:
\be520
\ba{rcl}
0= \delta \psi_{m i}
& =& {1 \over 2} \, e^A \Gamma_n \Gamma_5 \hat e_{\underline m}^{\ n}
    \Big[ e^{-A}\, \lambda \, \Theta \cdot Q
      +    \dot A \, {\mathbb 1} + P \cdot \Theta
    \Big]_i^{\ j} \epsilon_j \ ,
\ea
\ee
($P \equiv P^x \sigma^x$) which vanishes if
\be530
-(P^x  + \lambda e^{-A} Q^x ) \Theta^x
    = \dot A  \qquad , \qquad
\epsilon^{xyz} (P^x  +  \lambda e^{-A} Q^x) \Theta^y = 0 \ .
\ee
An obvious solution for $\Theta^x$ is
\be532
\Theta^x = \pm  {P^x  + \lambda \, e^{-A} Q^x \over
|P^x  + \lambda e^{-A} \, Q^x|}
 \qquad , \qquad \dot A = \mp |P^x  + \lambda e^{-A} Q^x | \ .
\ee
The second equation is the modified flow equation for the warp factor in
the metric and we infer that for flat walls ($\lambda=0$) the phase in
the projector coincides with the phase of the gravitino mass matrix,
i.e.\ $\Theta^x = P^x/|P^x|$. Non-zero curvature is related to a
deviation of the two phases.

Before we address the transversal (radial) component of the gravitino
variation, let us discuss the hyperino variation, which yields the flow
equation for the scalar fields. In order to obtain the flow equation we
use the projector and write
\be540
0= V^i_{u \alpha} \Gamma^5 \Big[ \dot q^u \, {\mathbb 1}  +
    \, 3 \, k^u \Theta\Big]_i^{\ j} \epsilon_j \ .
\ee
Multiplying this equation with the quaternionic vielbein $V^{k\ \alpha}_{
v}$ and using eq.\ (\ref{940}) yields after some steps the new flow equation
for the scalar fields
\be620
h_{uv} \dot q^v = 3 \, \Theta^x (\nabla_u P)^x \ .
\ee

Finally, from the radial component of the gravitino variation we
derive a constraint that fixes the phase $Q^x$. Again by employing
the projector this variation becomes
\be533
\ba{l}
\delta \psi_z \sim
    \Big[ \partial_z  + Q_z
    + {1 \over 2} P \cdot \Theta  \Big]_i^{\ j} \epsilon_j\ .
\ea
\ee
Since we are dealing here with systems of ordinary differential equations
the integration of these equations should at least in principle always be
possible yielding an expression for the Killing spinor. But this solution
has to be consistent with the projector (\ref{490}), which is not
obvious. Namely, we can also write $\nabla_z \epsilon \equiv
(\partial_z+Q_z) \epsilon = -(\partial_z+Q_z) \Gamma^5 \Theta \,
\epsilon$ and find an alternative expression
\be537
\delta \psi_z \sim \Gamma^5 \Big[-\nabla_z  \Theta
    - {1 \over 2} P \Big]_i^{\ j} \epsilon_j\ .
\ee
Multiplying this equation by $\Theta$ ($\Theta^2 = {\mathbb 1}$),
dropping $\Gamma^5$ and adding this equation to (\ref{533}) gives:
\be543
 \Theta \nabla_z \Theta = {1 \over 2}
    [\Theta, \nabla_z \Theta]= - {1\over 2} \, [\Theta,P] \ .
\ee
For flat walls with $\Theta = P/|P|$ one obtains the constraint $dq^u [
P, \nabla_u P]=0$, which was mentioned in the paper \cite{020}, but which
is fulfilled if the relation (\ref{580}) holds. Using the expression for
$\Theta$ in (\ref{532}) and since $\nabla_z P^x \sim \Theta^x$, see
(\ref{580}), combined with the flow equation $\nabla_z A = \mp |P^x +
\lambda e^{-A} Q^x|$ this constraint can be written as
\be116
\lambda \, e^{-A} [\Theta,  \nabla_z A ] =0\ .
\ee
For $\lambda \neq 0$ the solution becomes $\nabla_z A \sim \Theta$ or
in components:
\be837
\nabla_z Q^x = B \,  \Theta^x \ .
\ee
These are three differential equations which determine the two
non-trivial components of $Q^x$ (recall $Q^x Q^x = 1$) and the scalar
function $B$. In the next section we will investigate this relation
further.


\section{Solving the equations of motion}


Solutions of the the Killing spinor equations do not automatically
satisfy also the equations of motion. For black holes e.g., one
has to impose the Bianchi identities; see also the discussion in
\cite{040,050}. So let us verify that our flow equations
 (\ref{532}), (\ref{620}) combined with constraint (\ref{837}) solve the
equations of motion. Before we start with the explicit equations we will
derive some useful identities.

First, using the definition (\ref{284}) for the Killing prepotentials we
can write $\dot q^u$ also as
\be662
\dot q^u = 3 h^{uv} \Theta^x \nabla_v P^x = -
    3 \, h^{uv} \Omega_{vr}^x k^r \Theta^x \ .
\ee
One can contract this relation with the Killing vector $k_u$ and obtains
\be562
0= k_u \dot q^u = 3  \Big( k^u \nabla_u P^x \Big) \Theta^x\ ,
\ee
which ensures that it is consistent to decouple all gauge fields (see
discussion after equation (\ref{284})) and it implies that the flow
becomes perpendicular to the Killing direction. Moreover, using the
expression for $\dot q^u$ and (\ref{836}) and we derive
\be580
\nabla_z P^x = \nabla_v P^x \dot q^v =  k^w \Omega^x_{wv} \dot q^v =
3 |k|^2 \Theta^x \
,
\ee
After a  multiplication with $\Theta^x$ (\ref{580}) we find:
\be821
|k|^2 = {1 \over 3} \Theta^x (\nabla_z P)^x = { 1 \over 3} \Theta^x
\nabla_u P^x \dot q^u = (\Theta^x \nabla_u P^x)(\Theta^y \nabla^u P^y)
= {1 \over 9}\, |\dot q|^2 \ .
\ee
After having an expression for $\nabla_z P^x$ we can also derive
analogous expression for $\nabla_z Q^x$ and $\nabla_z \Theta^x$.
For this we  explore the constraint (\ref{837}) in more detail.
There are two possibilities, $B$ can vanish or not. In the first
case one can show that the equations of motion are not solved, see
also \cite{050}. If however $B\neq 0$ we infer from $Q^x \nabla_z
Q^x = 0$ that $\Theta^x Q^x = 0$ or $P^x Q^x = - \lambda e^{-A}$
and therefore the flow equation for the warp factor can be
simplified to
\be771
- \dot A = \pm \sqrt{P^2 - \lambda^2 e^{-2A}} = \Theta^x P^x
\ee
Next, the solution of eq.\  (\ref{543}) reads
\be829
\nabla_z \Theta^x = -P^x + C \Theta^x =  \lambda e^{-A} Q^x
\ee
where we used in the second step that $\Theta^x P^x = C$. Similarly we
find for the constraint (\ref{837})
\be736
\nabla_z Q^x = - \lambda e^{-A} \,  \Theta^x \ .
\ee
Having all covariant derivatives of the SU(2)-valued quantities we can
now start to discuss the equations of motion. The relevant Lagrangian has
the form:
\be163
S = \int \Big[ {R \over 2} - {1 \over 2} \, h_{uv} \partial q^u \partial q^v
    - V(q) \Big] \ ,
\ee
with the potential given by
\be364
V(q) = {3 \over 2} \Big( 3 \, |k|^2 - 4 \, P^x P^x \Big)
\ee
(see \cite{090} for the complete expressions).
With our metric Ansatz the Einstein equations become
\be937
\ba{l}
R_m^{\ n} = - \Big[ \ddot A +4 \dot A^2 + 3 \, e^{-2A} \lambda^2 \Big] \,
    \delta_m^{\ n} =  {2 \over 3} \, V \, \delta_m^{\ n}
    \ , \\[2mm]
R_z^{\ z} = -4 \Big[ \ddot A + \dot A^2 \Big] = h_{uv} \dot q^u \dot q^v
     + {2 \over 3} \, V \ .
\ea
\ee
Using our expressions from above we find for $\ddot A$:
\be663
\ba{rcl}
\ddot A &=& \mp {P^x \nabla_z P^x + \lambda^2 \dot A e^{-2A} \over
    \sqrt{P^2  - \lambda^2 e^{-2A}}} \\[2mm]
    &=& - 3 |k|^2 + \lambda^2 e^{-2A}
\ea
\ee
and it becomes straightforward to verify the Einstein equations (after
using the flow equations and eq.\ (\ref{821})). The scalar equation of
motion is given by
\be721
{1 \over \sqrt{g}} \partial_z \Big( \sqrt{g}\,  h_{uv} \dot q^v \Big)
- {1 \over 2} (\partial_u h_{rs}) \dot q^r \dot q^s =
\partial_u V
\ee
and we obtain for the different
components
\be992
\ba{rcl}
\partial_u V &=& 3 \dot q^v \Theta^x \nabla_u \nabla_v P^x -
    12 P^x \nabla_u P^x - {1 \over 2} \dot q^r (\partial_u h_{rs})
    \dot q^s \\[2mm]
{\partial_z \sqrt{g} \over \sqrt{g}} h_{uv} \dot q^v &=& -12 \Big(
    P^x \nabla_u P^x + \lambda e^{-A} Q^x \nabla_u P^x\Big) \\[2mm]
\partial_z (h_{uv} \dot q^v) &=&  3 \nabla_z (\Theta^x \nabla_u P^x )
    = 3 \Big( 4 \lambda e^{-A} Q^x \nabla_u P^x + \Theta^x \dot q^v
    \nabla_u \nabla_v P^x\Big)
\ea
\ee
where in the last step we used the relation $[\nabla_v , \nabla_u] P^x =
- \epsilon^{xyz} \Omega^y_{vu} P^z$ from \cite{110}. Thus all terms
cancel and we have shown that the flow equations solve the equations of
motion.

Note, we can integrate the equation for
the warp factor by introducing a new radial coordinate $y = y(z)$
defined by
\be923
{\partial y \over \partial z}   = \sqrt{|P|^2 - \lambda^2 \, e^{-2A}}\ ,
\ee
so that the equation for the warp factor becomes
$A' = \partial_y A = \mp 1$ and the new metric reads:
\be843
ds^2 = e^{\mp 2 (y - y_0) } \, \hat e^m \hat e^m + {dy^2 \over
    |P|^2 - \lambda^2 \, e^{\pm 2 (y - y_0)} } \ ,
\ee
where one has to insert into $|P|^2 = P^x P^x$ a solution of the
scalar flow equations, which in the new radial coordinate takes
the following form:
\be396
h_{uv} {dq^v \over dy} = \pm { \Theta^x \nabla_u P^x \over \Theta^x P^x}
\ .
\ee
{From} the metric we can conclude that this curved wall solution
can never asymptote to an IR critical point, where $e^{\mp 2y}
\rightarrow 0$ while $|P|^2$ stays finite. Also a flat space
vacuum with $|P|^2 \rightarrow 0$ while the warp factor stays finite is
cut-off. On the other hand, near an UV critical point where the warp
factor diverges ($e^{\mp 2y} \rightarrow \infty$) while $|P|^2 \neq 0$,
the effect of the wall curvature drops out. We cannot make general
statements about singular infra-red flows where $|P|^2$ and $e^{\pm 2 y}$
diverges.


\section{Example: Bent walls with the universal hypermultiplet}


In this Section we will discuss two concrete examples.  In both cases
we consider the coset $SU(2,1)/U(2)$ related to the universal
hypermultiplet, but in a different parameterization. In the model I
the metric is explicitly spherical symmetric so that some scalars are
compact, whereas in model II all scalars are non-compact.


\subsection{Model I}


We will use the notation used in \cite{020} and parameterize this coset
space in terms of two complex scalars $z_1$ and $z_2$ with the K\"ahler
potential:
\be164
K = - \log(1- |z_1|^2 - |z_2|^2 )\ ,
\ee
and the K\"ahler metric and the K\"ahler two-form take the form:
\be173
\ba{rcl}
\partial_A \partial_{\bar B} K\, dz^A d\bar z^B
&=& e^{K} \delta_{AB} \, dz^A d\bar z^B + e^{2K}
(\bar z_A dz^A)(z_B d\bar z^B)\, ,  \\
\partial_{A}\partial_{\bar B} K\,  dz^A \wedge d\bar z^B
&=& e^{K} \delta_{AB} \, dz^A \wedge d\bar z^B+ e^{2K}
 (\bar z_A \, dz^A) \wedge (z_B d\bar z^B) \ .
\ea
\ee
Following essentially the parameterization employed in
\cite{180}, it turns out to be more convenient to introduce polar
coordinates in the following way:
\be191
z_1 = r \, (\cos\theta/2) \, e^{i (\psi + \varphi)/2} \qquad , \qquad
z_2 = r \, (\sin\theta/2) \, e^{i (\psi - \varphi)/2}\, ,
\ee
with $r \in [0,1),\ \theta \in [0, \pi),\ \varphi \in [0, 2 \pi)$
and $\psi \in [0, 4\pi)$. The K\"ahler metric becomes:
\be202
\partial_A \partial_{\bar B} K\, dz^A d\bar z^B
= {dr^2 \over (1- r^2 )^2} + {r^2 \over 4 (1- r^2 )}
(\sigma_1^2 + \sigma_2^2 ) +{r^2 \over 4 (1- r^2 )^2} \sigma_3^2 \ ,
\ee
where the $SU(2)$ one-forms ($d\sigma_i +{1\over 2}
\epsilon_{ijk} \sigma_j \wedge \sigma_k =0$) are given by:
\be211
\ba{l}
\sigma_1 = \cos\psi \, d\theta + \sin\psi \, \sin\theta\, d\varphi \ , \\
\sigma_2 = - \sin\psi \, d\theta + \cos\psi \, \sin\theta \, d\varphi \ , \\
\sigma_3 = d\psi + \cos\theta \, d\varphi  \ .
\ea
\ee
For this model one obtains for the SU(2) connection is given \cite{020}
\be253
\omega^1 = - {\sigma_1 \over \sqrt{1-r^2}} \quad , \quad
\omega^2 =  {\sigma_2 \over \sqrt{1-r^2}} \quad , \quad
\omega^3 = -{1 \over 2}(1 + {1 \over 1-r^2}) \, \sigma_3 \ .
\ee
In order to keep the Killing prepotential as simple as possible we
gauge the Killing vector $ k = \partial_{\psi}$ and the corresponding
Killing prepotential becomes
\be601
P^x = {r^2 \over 2( 1-r^2)} \, \delta^{x3} \ .
\ee
The general form of all Killing prepotentials can found in \cite{020}.
Using (\ref{771}) and the fact that the SU(2) connection (\ref{253})
have no radial component, the radial flow equation for the upper sign
becomes
\be722
g_{rr} r'(y) = 3 {\Theta^3 \partial_r P^3 \over \sqrt{|P|^2 - \lambda^2
e^{-2A}}} = 3 \, \partial_r \log P^3
\ee
which is solved by
\be782
r^2 = 1- e^{-12 y} \ ,
\ee
$y \in (0, +\infty)$. Therefore the metric reads
\be766
ds^2 = e^{-2 y} \, \hat e^m \hat e^m + { 4 dy^2 \over (e^{12y} -1)^2 -
    4 \lambda^2 e^{2y}} \ .
\ee
But this is not the complete solution, there is a second scalar
running and we have to investigate the remaining equations
\be665
h_{ij} (q^j)' = 3 {\Theta^x \nabla_i P^x \over \Theta^3 P^3}
    = 3 \lambda e^{-A}(Q^1 \omega^2_i - Q^2 \omega^1_i) {P^3
    \over |P|^2 - \lambda^2 e^{-2A}}
\ee
Since $\omega^{1/2}$ have no $\psi$-component we find that
$h_{\psi j} (q^j)' = 0$, which can be solved by $\psi' =
\varphi'=0$. This implies that $h_{\varphi j} (q^j)' = 0$, which is
the case if: $Q^1 \cos \psi + Q^2 \sin\psi = 0$. Together with
constraint $|Q|=1$ this implies
\be652
Q^2 =- \cot\psi \, Q^1 = {\sqrt{|P|^2 - \lambda^2 e^{-2A}}
    \over P^3} \, \cos\psi
\ee
(recall $\psi$ is constant). Using this relation in the equation $h_{\theta j}
(q^j)' = 0$ gives
\be362
h_{\theta\theta} \, \theta' = {3 \lambda e^{-A}  \over
    \sqrt{1-r^2} \sqrt{|P|^2 - \lambda^2 e^{-2A}}}
\ee
and after inserting $P^3$ and the solution for $r^2$ in
(\ref{782}) we get the differential  whose solution is of the
form:
\be621 \theta(y)  = \int^y{ 24 \, \lambda \, e^{7 y'} \over
(e^{12y'}-1) \sqrt{
    (e^{12y'}-1)^2 - 4 \lambda^2 e^{2y'}}} \ dy'\ .
\ee
Eventhough the above integral cannot be expressed in terms of
known functions, one can infer that at the critical value of the
$y$ component the field $\theta$ remains finite.

Note that  we have determined the phase $Q$ in (\ref{652}) from
the flow equation, but this phase has to solve the differential
equation (\ref{736}). We have checked explicitly that this
nontrivial constraint is indeed satisfied.

In summary, we considered here a curved wall generalization of a
flow to flat spacetime (with $|P|^2=0$). The solution shows that
if one turns on the wall curvature ($\lambda \neq 0$), the flat
space vacuum at $r=0$ is cutoff and the flows stops at some
finite value of $r$ where the transversal metric component has a
pole, however the space-time remains regaular.  Our calculations
also show, that in contrast to the flat walls the curved wall
requires an additional nontrivial scalar, whereas the remaining
two are arbitrary constants.


\subsection{Model II}


In the second model we parameterize the same coset space in a
different way by using the K\"ahler potential \cite{300}
\be726
K =-{1 \over 2} \log( S + \bar S - 2 C\bar C)
\ee
and write: $S= V + \theta^2 + \tau^2 + i \sigma$
and $C = \theta - i \tau$ which yield the metric
\be355
ds^2 = {dV^2 \over 2 V^2} +{1 \over 2 V^2} \Big[ d\sigma +
    2(\theta d\tau -\tau d\theta)\Big]^2 + {2 \over V}
    \Big[ d\tau^2 + d\theta^2 \Big]
\ee
and the SU(2) connection becomes
\be246
\omega^1 = -{d\tau \over \sqrt{V}} \ , \quad
\omega^2 = {d\theta \over \sqrt{V}} \ , \quad
\omega^3 = -{1 \over 4 V} \Big[d\sigma + 2( \theta d\tau - \tau
d\theta)\Big] \ .
\ee
We took these expression from \cite{080} (see also \cite{310}), where
a different convention is used, which basically implies an additional
factor of ``2'' in the covariant derivatives.

As before we choose a gauging where the Killing prepotential
becomes simple, which is the case for $k = \partial_{\sigma}$
so that
\be994
P^x = - {1 \over 4 V}\, \delta^{x3} \ .
\ee
The computation goes analogous to the Model I. The connection $\omega^x$
has no $V$-component and hence
\be732
g_{VV} V'(y) = 3 {\Theta^3 \partial_V P^3 \over \sqrt{|P|^2 - \lambda^2
e^{-2A}}} = 3 \, \partial_V \log P^3
\ee
which is solved by
\be723
V = e^{-6y}
\ee
where we again dropped the integration constant. Thus, the metric becomes
\be121
ds^2 = e^{-2 y} \, \hat e^m \hat e^m + { 16 \, dy^2 \over e^{12y} -
    16 \, \lambda^2 e^{2y}}
\ee
To solve the remaining equations note that $\omega^{1/2}$ have no
$\sigma$-component and hence $\Theta^x \nabla_{\sigma} P^x \sim (\Theta^1
\omega^2_{\sigma} - \Theta^2 \omega^1_{\sigma}) =0$
yielding $h_{\sigma i} \dot q^i = 0$. This equations is solved if
\be729
\theta = c\, \tau \qquad , \qquad \sigma = constant
\ee
for some constant  $c$. {From} this relation follows that $h_{\theta
i} \dot q^i - c \, h_{\tau i} \dot q^i = 0$ which yields
\be524
Q^1 = c\, Q^2 = {c \, \sqrt{|P|^2 - \lambda^2 e^{-2A}} \over
    \sqrt{1 +c^2} \; P^3}
\ee
and we find as differential equation for $\tau$
\be892
\tau' = -{3 \, c \, \lambda \over \sqrt{1 + c^2}} \, {\sqrt V e^{-A} \over
    \sqrt{{1\over 16 V^2} -\lambda^2 e^{-2A}}}
= -{12 c \, \lambda \over \sqrt{1 + c^2}} \, {e^{-2y} \over
    \sqrt{e^{12y} - 16 \lambda^2 e^{2y}}}
\ee
which can be solved explicitly
\begin{equation}
\theta = c\, \tau = 3\,\sqrt{{\displaystyle \frac {A}{1 + c^{2}}} }\,
\left(  \!
{\displaystyle \frac {1}{2}} \,{\displaystyle \frac {x}{A\,(x^{2}
 + A)^{(1/5)}}}  - {\displaystyle \frac {3}{10}} \,
{\displaystyle \frac {x\,\mathrm{hypergeom}\Big([{\displaystyle \frac
{1}{2}} , \,{\displaystyle \frac {1}{5}} ], \,[ {\displaystyle
\frac {3}{2}} ], \, - {\displaystyle \frac {x^{2} }{A}}
\Big)}{A^{(6/5)}}}  \!  \right)
\end{equation}
where $A=16\lambda^2$ and $x=\sqrt{e^{10 y}-A}$. Again at the
critical  value of $y$, $\theta$ remains finite.

As before, we have to check that the phase $Q$ in (\ref{524})
satisfies the equation (\ref{736}), which is again the case. Also
in this example there are two scalars that flow, whereas the
other two remain constant.  In the flat wall limit ($\lambda =
0$) this model has a known M-theory embedding and corresponds to
the intersection of three 5-branes over a common 3-brane, which
becomes the domain wall upon compactification.  This is a
well-known supergravity solution that has been discussed in more
detail in \cite{310, 320} (and refs.\ therein).  In this setup
the scalar $V$ is basically the volume of internal space, whereas
the scalars $\tau$ and $\theta$ are related to the radii of the
(3,0) and (0,3) cycles. Therefore, a non-trivial wall curvature
gives a cutoff for the volume scalar $V$ where the tranversal
metric component develops a pole. At this point the scalars
$\tau$ and $\theta$ are driven to zero and we expect a
singularity in the internal space.


\bigskip

\bigskip

{\bf Acknowledgments}

\nopagebreak

We would like to thank Gianguido Dall'Agata and Gabriel Lopes
Cardoso  and  Neil Lambert for discussions. The work is supported
by a DFG Heisenberg Fellowship (K.B.), the U.S. Department of
Energy Grant No. DOE-EY-76-02-3071 (M.C.) and the NATO Linkage
grant No. 97061  and Class of 1965 Endowed Term Chair(M.C.).  The
authors  would like to thank UPenn High Energy Theory group
(K.B.), High Energy Theory group at Rutgers University  (M.C.) and
the Institute for Theoretical Physics at the UCSB (K.B. and M.C.)
for hospitality during different stages of the work.

\vspace{10mm}




\appendix{\bf \Large Appendix}
\medskip

{\bf \large A. Conventions}

\bigskip

The SU(2) indices are raised with the $\epsilon$-tensor
\be900
\epsilon_i = \epsilon_{ij} \epsilon^j
\qquad , \qquad
\epsilon^i = \epsilon^{ji} \epsilon_j
\ee
with $\epsilon^{12} = \epsilon_{12} =1$ and similarly the
$Sp(n)$ indices with
\be910
V_{\alpha} = C_{\alpha\beta} V^{\beta}
\qquad , \qquad
V^{\alpha} = C^{\beta\alpha} V_{\beta}
\ee
with $C_{\alpha\beta} = - C_{\beta\alpha}$ and $C^2 = - {\mathbb 1}$.
The three complex structures fulfill the algebra
\be930
J^x \cdot J^y = - \delta^{xy} \, {\mathbb 1} + \epsilon^{xyz} J^z
\ee
and because the SU(2) curvatures are given by
$\Omega^x = e^m J^x_{mn} \wedge e^n$
this relation becomes
\be836
h^{vt} \Omega_{uv}^x \Omega_{tw}^y = - \delta^{xy} h_{uw}
    + \epsilon^{xyz} \Omega^z_{uw} \ .
\ee
For the quaternionic vielbeine holds the relation
\be940
2\, V_{u \alpha}^i V_{v}^{j\  \alpha} = h_{uv} \epsilon^{ij}
+ i  (\Omega_{uv})^{ij}
\ee
with $(\Omega_{uv})^{ij} = i \, \Omega^x_{uv} (\sigma^x)_k^{\ j} \epsilon^{ki}$
and since the Pauli matrices are traceless one find $(\Omega_{uv})^{ij}
\epsilon_{ij} =0$.

\bigskip




\begin{thebibliography}{10}

\bibitem{cgr}
M.~Cveti\v c, S.~Griffies, and S.-J. Rey, ``Static domain walls
in {N=1}
  supergravity,'' {\em Nucl. Phys.} {\bf B381} (1992) 301--328,
  \href{http://xxx.lanl.gov/abs/http://arXiv.org/abs/hep-th/9201007}{{\tt
  http://arXiv.org/abs/hep-th/9201007}}.

\bibitem{c}
M.~Cveti\v c, ``Extreme domain wall - black hole complementarity
in {N=1}
  supergravity with a general dilaton coupling,'' {\em Phys. Lett.} {\bf B341}
  (1994) 160--165.

\bibitem{cs}
M.~Cveti\v c and H.~H. Soleng, ``Supergravity domain walls,''
{\em Phys. Rept.}
  {\bf 282} (1997) 159--223,
  \href{http://xxx.lanl.gov/abs/http://arXiv.org/abs/hep-th/9604090}{{\tt
  http://arXiv.org/abs/hep-th/9604090}}.

\bibitem{clp}
M.~Cveti\v c, H.~Lu, and C.~N. Pope, ``Consistent {Kaluza-Klein}
sphere
  reductions,'' {\em Phys. Rev.} {\bf D62} (2000) 064028,
  \href{http://xxx.lanl.gov/abs/http://arXiv.org/abs/hep-th/0003286}{{\tt
  http://arXiv.org/abs/hep-th/0003286}}.

\bibitem{250}
D.~Z. Freedman, S.~S. Gubser, K.~Pilch, and N.~P. Warner, ``Renormalization
  group flows from holography supersymmetry and a c-theorem,'' {\em Adv. Theor.
  Math. Phys.} {\bf 3} (1999) 363--417,
  \href{http://xxx.lanl.gov/abs/hep-th/9904017}{{\tt hep-th/9904017}}.

\bibitem{080}
A.~Ceresole, G.~Dall'Agata, R.~Kallosh, and A.~Van~Proeyen, ``Hypermultiplets,
  domain walls and supersymmetric attractors,''
  \href{http://xxx.lanl.gov/abs/hep-th/0104056}{{\tt hep-th/0104056}}.

\bibitem{rs}
L.~Randall and R.~Sundrum, ``An alternative to compactification,'' {\em Phys.
  Rev. Lett.} {\bf 83} (1999) 4690--4693,
  \href{http://xxx.lanl.gov/abs/http://arXiv.org/abs/hep-th/9906064}{{\tt
  http://arXiv.org/abs/hep-th/9906064}}.

\bibitem{cg}
M.~Cveti\v c and S.~Griffies, ``Gravitational effects in
supersymmetric domain
  wall backgrounds,'' {\em Phys. Lett.} {\bf B285} (1992) 27--34,
  \href{http://xxx.lanl.gov/abs/http://arXiv.org/abs/hep-th/9204031}{{\tt
  http://arXiv.org/abs/hep-th/9204031}}.

\bibitem{bcy}
F.~Brito, M.~Cveti\v c, and S.~Yoon, ``From a thick to a thin
supergravity domain
  wall,'' {\em Phys. Rev.} {\bf D64} (2001) 064021,
  \href{http://xxx.lanl.gov/abs/http://arXiv.org/abs/hep-ph/0105010}{{\tt
  http://arXiv.org/abs/hep-ph/0105010}}.

\bibitem{150}
R.~Kallosh and A.~Linde, ``Supersymmetry and the brane world,'' {\em JHEP} {\bf
  02} (2000) 005, \href{http://xxx.lanl.gov/abs/hep-th/0001071}{{\tt
  hep-th/0001071}}.

\bibitem{140}
K.~Behrndt and M.~Cveti\v c, ``Anti-de sitter vacua of gauged
supergravities with
  8 supercharges,'' {\em Phys. Rev.} {\bf D61} (2000) 101901,
  \href{http://xxx.lanl.gov/abs/hep-th/0001159}{{\tt hep-th/0001159}}.

\bibitem{al/co}
D.~V. Alekseevsky, V.~Cortes, C.~Devchand, and A.~Van~Proeyen, ``Flows on
  {quaternionic-Kaehler} and very special real manifolds,''
  \href{http://xxx.lanl.gov/abs/http://arXiv.org/abs/hep-th/0109094}{{\tt
  http://arXiv.org/abs/hep-th/0109094}}.

\bibitem{270}
K.~Behrndt and G.~Dall'Agata, ``Vacua of {N} = 2 gauged supergravity derived
  from non- homogenous quaternionic spaces,''
  \href{http://xxx.lanl.gov/abs/http://arXiv.org/abs/hep-th/0112136}{{\tt
  http://arXiv.org/abs/hep-th/0112136}}.

\bibitem{cgs}
M.~Cveti\v c, S.~Griffies, and H.~H. Soleng, ``Local and global
gravitational
  aspects of domain wall space-times,'' {\em Phys. Rev.} {\bf D48} (1993)
  2613--2634,
  \href{http://xxx.lanl.gov/abs/http://arXiv.org/abs/gr-qc/9306005}{{\tt
  http://arXiv.org/abs/gr-qc/9306005}}.

\bibitem{060}
O.~DeWolfe, D.~Z. Freedman, S.~S. Gubser, and A.~Karch, ``Modeling the fifth
  dimension with scalars and gravity,'' {\em Phys. Rev.} {\bf D62} (2000)
  046008, \href{http://xxx.lanl.gov/abs/hep-th/9909134}{{\tt hep-th/9909134}}.

\bibitem{cw}
M.~Cveti\v c and J.~Wang, ``Vacuum domain walls in
{D-dimensions}: {Local} and
  global space-time structure,'' {\em Phys. Rev.} {\bf D61} (2000) 124020,
  \href{http://xxx.lanl.gov/abs/http://arXiv.org/abs/hep-th/9912187}{{\tt
  http://arXiv.org/abs/hep-th/9912187}}.

\bibitem{210}
A.~Karch and L.~J. Randall, ``Locally localized gravity,'' {\em JHEP} {\bf 05}
  (2001) 008, \href{http://xxx.lanl.gov/abs/hep-th/0011156}{{\tt
  hep-th/0011156}}.

\bibitem{090}
A.~Ceresole and G.~Dall'Agata, ``General matter coupled {N=2,D=5} gauged
  supergravity,'' {\em Nucl. Phys.} {\bf B585} (2000) 143--170,
  \href{http://xxx.lanl.gov/abs/hep-th/0004111}{{\tt hep-th/0004111}}.

\bibitem{170}
M.~Gunaydin and M.~Zagermann, ``The vacua of 5d, {N} = 2 gauged
  {Yang-Mills/Einstein/tensor} supergravity: {Abelian} case,'' {\em Phys. Rev.}
  {\bf D62} (2000) 044028, \href{http://xxx.lanl.gov/abs/hep-th/0002228}{{\tt
  hep-th/0002228}}.

\bibitem{100}
L.~Andrianopoli {\em et.~al.}, ``N = 2 supergravity and {N=2} super
  {Yang-Mills} theory on general scalar manifolds: {Symplectic} covariance,
  gaugings and the momentum map,'' {\em J. Geom. Phys.} {\bf 23} (1997)
  111--189, \href{http://xxx.lanl.gov/abs/hep-th/9605032}{{\tt
  hep-th/9605032}}.

\bibitem{030}
K.~Behrndt and M.~Cveti\v c, ``Supersymmetric domain wall world
from {D} = 5
  simple gauged supergravity,'' {\em Phys. Lett.} {\bf B475} (2000) 253,
  \href{http://xxx.lanl.gov/abs/hep-th/9909058}{{\tt hep-th/9909058}}.

\bibitem{be/he}
K.~Behrndt, C.~Herrmann, J.~Louis, and S.~Thomas, ``Domain walls in five
  dimensional supergravity with non- trivial hypermultiplets,'' {\em JHEP} {\bf
  01} (2001) 011,
  \href{http://xxx.lanl.gov/abs/http://arXiv.org/abs/hep-th/0008112}{{\tt
  http://arXiv.org/abs/hep-th/0008112}}.

\bibitem{010}
K.~Behrndt, G.~L. Cardoso, and D.~{L\"ust}, ``Curved {BPS} domain wall
  solutions in four-dimensional {N} = 2 supergravity,''
  \href{http://xxx.lanl.gov/abs/hep-th/0102128}{{\tt hep-th/0102128}}.

\bibitem{070}
G.~Lopes~Cardoso, G.~Dall'Agata, and D.~{L\"ust}, ``Curved {BPS} domain wall
  solutions in five-dimensional gauged supergravity,''
  \href{http://xxx.lanl.gov/abs/hep-th/0104156}{{\tt hep-th/0104156}}.

\bibitem{040}
A.~H. Chamseddine and W.~A. Sabra, ``Curved domain walls of five dimensional
  gauged supergravity,'' \href{http://xxx.lanl.gov/abs/hep-th/0105207}{{\tt
  hep-th/0105207}}.

\bibitem{050}
A.~H. Chamseddine and W.~A. Sabra, ``Einstein brane-worlds in {5D} gauged
  supergravity,'' \href{http://xxx.lanl.gov/abs/hep-th/0106092}{{\tt
  hep-th/0106092}}.

\bibitem{160}
J.~Bagger and E.~Witten, ``Matter couplings in {N=2} supergravity,'' {\em Nucl.
  Phys.} {\bf B222} (1983) 1.

\bibitem{110}
R.~D'Auria and S.~Ferrara, ``On fermion masses, gradient flows and potential in
  supersymmetric theories,'' {\em JHEP} {\bf 05} (2001) 034,
  \href{http://xxx.lanl.gov/abs/hep-th/0103153}{{\tt hep-th/0103153}}.

\bibitem{020}
K.~Behrndt and M.~Cveti\v c, ``Gauging of {N=2} supergravity
hypermultiplet and
  novel renormalization group flows,''
  \href{http://xxx.lanl.gov/abs/hep-th/0101007}{{\tt hep-th/0101007}}.

\bibitem{180}
R.~Britto-Pacumio, A.~Strominger, and A.~Volovich, ``Holography for coset
  spaces,'' {\em JHEP} {\bf 11} (1999) 013,
  \href{http://xxx.lanl.gov/abs/hep-th/9905211}{{\tt hep-th/9905211}}.

\bibitem{300}
S.~Ferrara and S.~Sabharwal, ``Quaternionic manifolds for type II
superstring
  vacua of calabi-yau spaces,'' {\em Nucl. Phys.} {\bf B332} (1990) 317.

\bibitem{310}
A.~Lukas, B.~A. Ovrut, K.~S. Stelle, and D.~Waldram, ``Heterotic M
-theory in
  five dimensions,'' {\em Nucl. Phys.} {\bf B552} (1999) 246--290,
  \href{http://xxx.lanl.gov/abs/http://arXiv.org/abs/hep-th/9806051}{{\tt
  http://arXiv.org/abs/hep-th/9806051}}.

\bibitem{320}
K.~Behrndt and S.~Gukov, ``Domain walls and superpotentials from
M theory on
  Calabi- Yau three-folds,'' {\em Nucl. Phys.} {\bf B580} (2000) 225--242,
  \href{http://xxx.lanl.gov/abs/http://arXiv.org/abs/hep-th/0001082}{{\tt
  http://arXiv.org/abs/hep-th/0001082}}.

\end{thebibliography}

\providecommand{\href}[2]{#2}\begingroup\raggedright\endgroup

\end{document}